\documentclass{sig-alternate}

\conferenceinfo{KDD}{'09 Paris, France}

\newcommand{\figref}[1]{Fig.~#1}

\newcommand{\secref}[1]{Sec.~#1}

\newcommand{\panela}{A}
\newcommand{\panelb}{B}
\newcommand{\panelc}{C}
\newcommand{\paneld}{D}
\newcommand{\panele}{E}

\newcommand{\paneli}{I}

\newcommand{\nhpp}{non-homogeneous Poisson process~}
\newcommand{\cnhpp}{cascading non-homogeneous Poisson process~}

\newcommand{\cnhppnospace}{cascading non-homogeneous Poisson process}

\newcommand{\zzz}{z}      
\newcommand{\zn}[1]{\zzz_{#1}}      
\newcommand{\Z}{{\boldsymbol \zzz}}               
\newcommand{\T}{{\boldsymbol t}}               
\newcommand{\ttt}{t}
\newcommand{\deltat}{\Delta \ttt}

\newcommand{\tn}[1]{\ttt_{#1}}               
\newcommand{\deltatn}[1]{\deltat_{#1}}
\newcommand{\n}{n}                        
\newcommand{\N}{N}                        
\newcommand{\p}{p}                         

\newcommand{\bftheta}{{\boldsymbol \theta}}

\newcommand{\qq}{\xi}
\newcommand{\rate}{\rho}
\newcommand{\ratet}[1]{\rate(#1)}
\newcommand{\ratea}{\rate_a}

\newcommand{\condprobthree}[3]{\p_{#1}(#2\,|\:#3)} 
\newcommand{\prob}[1]{\p(#1)}            
\newcommand{\probtwo}[2]{\p_{#1}(#2)}

\newcommand{\tauz}{\tau_0}
\newcommand{\tauo}{\tau_1}
\newcommand{\tauzd}{\tau_{0d}}
\newcommand{\tauod}{\tau_{1d}}
\newcommand{\tauzw}{\tau_{0w}}
\newcommand{\tauow}{\tau_{1w}}
\newcommand{\tauepsd}{\tauzd,\tauod,\varepsilon_d}
\newcommand{\tauepsw}{\tauzw,\tauow,\varepsilon_w}
\newcommand{\pulse}[3]{\prob{t;#1,#2,#3}}
\newcommand{\pdfull}{\condprobthree{d}{\ttt}{\tauepsd}}
\newcommand{\pwfull}{\condprobthree{w}{\ttt}{\tauepsw}}
\newcommand{\pd}{\probtwo{d}{\ttt}}
\newcommand{\pw}{\probtwo{w}{\ttt}}
\newcommand{\Nw}{\rate_p}
\newcommand{\ratei}{\Nw}
\newcommand{\ratep}{\Nw} 

\newcommand{\paramslist}{\ratei,\ratea,\qq,\tauepsd,\tauepsw}

\newcommand{\bfz}{\Z}
\newcommand{\bft}{\T}

\begin{document}

\title{Characterizing Individual Communication Patterns

}

\numberofauthors{1}
\author{
\alignauthor R.~Dean Malmgren$^{1,2}$, Jake M.~Hofman$^{1}$, 
Lu\'is A.~N.~Amaral$^{2}$, and Duncan J.~Watts$^{1}$ \\
\end{tabular}
\begin{tabular}{cc}
\affaddr $^1$Yahoo!~Research &
\affaddr $^2$Chemical \& Biological Engineering Department \\
\affaddr 111 W 40th St, 17th Floor & \affaddr Northwestern University \\
\affaddr New York, NY 10018, USA & \affaddr Evanston, IL 60208, USA\\
\eaddfnt \{hofman, djw\}@yahoo-inc.com & 
\eaddfnt \{r-malmgren, amaral\}@northwestern.edu \\
}


\date{2009 February 6}
\maketitle

\begin{abstract}
The increasing availability of electronic communication data, such as
that arising from e-mail exchange, presents social and information
scientists with new possibilities for characterizing individual
behavior and, by extension, identifying latent structure in human
populations.  Here, we propose a model of individual e-mail
communication that is sufficiently rich to capture meaningful
variability across individuals, while remaining simple enough to be
interpretable.  We show that the model, a cascading non-homogeneous
Poisson process, can be formulated as a double-chain hidden Markov
model, allowing us to use an efficient inference algorithm to estimate
the model parameters from observed data. We then apply this model to
two e-mail data sets consisting of 404 and 6,164 users, respectively,
that were collected from two universities in different countries and
years.  We find that the resulting best-estimate parameter
distributions for both data sets are surprisingly similar, indicating
that at least some features of communication dynamics generalize
beyond specific contexts.  We also find that variability of individual
behavior over time is significantly less than variability across the
population, suggesting that individuals can be classified into
persistent ``types''.  We conclude that communication patterns may
prove useful as an additional class of attribute data, complementing
demographic and network data, for user classification and outlier
detection---a point that we illustrate with an interpretable 
clustering of users based on their inferred model parameters.

\end{abstract}

\category{ \\I.6.1}{Simulations and Modeling}{Simulation theory}[model
  classification]



\section{Introduction}
\label{sec:introduction}

Empirical social science has long sought to classify humans according to their
individual attributes, whether observable or self-reported, and to construct
models based on these attributes that predict or explain other sociological
variables of interest, like group membership~\cite{mcpherson87}, political
preferences~\cite{lazarsfeld68}, or consumer
behavior~\cite{katz55}. Traditionally, this modeling exercise has relied on the
measurement of demographic attributes~\cite{hinde98}, like gender, age, race,
education and income, that are typically obtained through observational and
survey methods. Although straightforward in principle, these methods are
generally expensive and time consuming to implement, especially if one wishes
to avoid problems associated with sampling bias and respondent
reliability~\cite{bernard84}.  More importantly, demographic attributes are
often poor predictors of many of the outcome variables---such as behaviors,
attitudes, and beliefs---that are of interest to social
scientists~\cite{granovetter85}.

The explosive growth of Internet-related communication over the past decade is
therefore of great interest from a social science perspective as it offers new
sources of data, such as e-mail~\cite{eckmann04,kossinets06}, instant
messaging~\cite{leskovec08}, telephone~\cite{gonzalez08} and social networking
records~\cite{kleinberg08}, that can be used to approximate sociological
relationships and behavior for very large populations in an efficient and
reliable manner.  Until now, these data have mostly been used to construct
social networks, where individuals can then be characterized in terms of
structural attributes like degree, centrality, structural equivalence, and
community membership~\cite{newman03,wasserman94,sales-pardo07}.  Typically
these structural features are then related to other kinds of individual
characteristics, including demographic attributes, organizational affiliations,
and outcome variables like social capital~\cite{coleman88,portes98} or career
success~\cite{burt04}.

In this paper we introduce an alternative approach to characterizing
individuals that, like network analysis, takes activity data as input, but
focuses instead on the dynamics of the activity rather than the structure of
links with others.  In order to illustrate the approach, we study e-mail
communication activity---that is, we model the time series of outgoing messages
recorded by an individual's e-mail server---and we also propose a particular
model that captures some characteristic features of human communication
behavior.  Significantly, this approach to modeling e-mail data may find useful
applications in areas like spam detection, where the abusive behavior of
interest may exhibit temporal patterns that are characteristically
``non-human,'' or may simply be distinguishable from the previous behavior of a
given user. However, we emphasize at the outset that neither e-mail data nor
the particular model we specify is essential to the approach followed here.
Indeed our approach can be applied to other modes of communication, or even
other forms of activity, like web-browsing or clickstream data, that generate
time series of an individual's discrete actions. 
We emphasize that temporal patterns in individual behavior serve as an
additional set of features, analogous to demographic or network attributes,
that may be leveraged either to cluster users into groups of similar
individuals or to identify changes in behavior over time.

\textbf{Related Work.} Clearly, modeling of temporal data has been investigated
extensively, especially in the context of clustering time series~\cite{liao05},
where standard clustering techniques, such as K-means, agglomerative
hierarchical clustering or mixture model clustering have been used to
identify similarities among parameter estimates for models across a
wide range of disciplines~\cite{liao05}.  To the best of our
knowledge, however, these techniques have not been used specifically
to characterize human correspondence patterns or, more generally,
human activity patterns.  Thus although we also perform some
rudimentary clustering of our model parameters, our work should be
viewed as complementary to the clustering literature in that our focus
is on the formulation and estimation of the model itself, rather than
on the techniques used to cluster the resulting parameters.

In more closely related work, several models have recently been proposed
specifically to account for empirical regularities in human activity patterns.
One class of models in this genre asserts that human communication behavior can
be captured by simple, universally applicable rules, where individuals are
assumed to decide when to execute tasks based on a priority
queue~\cite{barabasi05,vazquez06a}. These models have been shown to reproduce
some asymptotic statistical properties of communication patterns; however, they
also make predictions that are clearly at odds with important features of the
empirical data~\cite{stouffer06,malmgren08}. In particular, priority queuing
models fail to account for two important features of human communication
patterns: first, that individuals are influenced by daily and weekly cycles of
activity; and second, that individuals tend to send multiple messages during
``sessions'' of relatively high activity.

In response to these shortcomings, an alternative class of models has been
proposed that explicitly takes these features into account~\cite{malmgren08}.
One such model, a so-called \cnhppnospace, has been shown to be consistent with
at least one empirical data set~\cite{eckmann04}. Unfortunately, the complexity
of the model also makes the corresponding parameter estimations difficult to
interpret; thus it is a poor candidate for the particular purpose we have in
mind here, namely the characterization of individuals in terms of a relatively
small number of interpretable attribute variables. On a practical note,
moreover, the relatively complex structure of the model proposed in
Ref.~\cite{malmgren08} requires expensive optimization techniques
(e.g. simulated annealing) to infer model parameters---a slow process that
renders the approach impractical for data sets larger than a few hundred
individuals.

\textbf{Our contributions.}
To resolve these issues, we propose a simplified parametric version of the
model in Ref.~\cite{malmgren08} that is sufficiently rich to differentiate
between individuals in a meaningful way, while remaining simple enough that we
can interpret these differences both in terms of individuals' daily and weekly
cycles, and also the extent to which they concentrate their effort into
sessions. In addition, we exploit the fact that a \cnhpp can be regarded as a
Markov mixture of Poisson processes, which can in turn be recast as a variant
of the canonical hidden Markov model~\cite{rabiner1989}.  We develop the
associated expectation-maximization algorithm for computationally efficient
parameter inference from the observed data, thereby rendering it practical for
estimating model parameters for a large population of users.

\begin{figure}
\epsfig{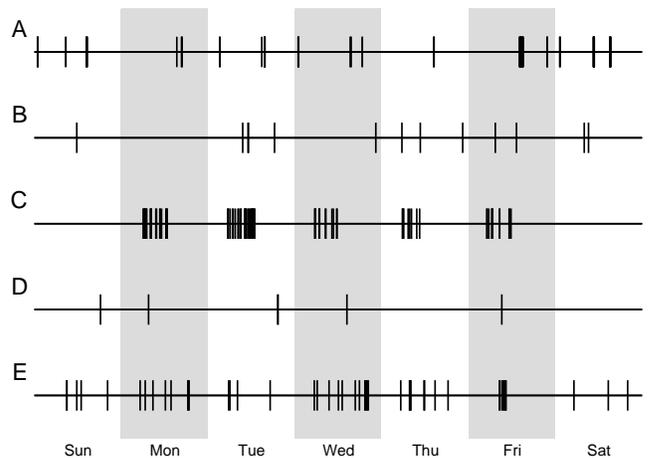}
\vspace*{-9pt}
\caption{
The challenges of characterizing human communication patterns.
{\bf \panela}--{\bf \panele}, Four empirical time series during a randomly
selected week and one synthetic time series generated from our \cnhpp with
unrealistic parameters for humans.
}
\vspace*{-9pt}
\label{fig:time_series}
\end{figure}

To demonstrate the utility of our approach, we apply this model to two
university e-mail data sets, and study the distributions of inferred
parameters, finding similar results both across universities and over
successive semesters. More importantly, we also show that individual parameter
estimates fluctuate less over time than they do across individuals, suggesting
that individual attributes are relatively persistent, and thus are reasonable
candidates for characterizing individuals. To illustrate this point, we refer
to the five time series in \figref{\ref{fig:time_series}}. Four of these time
series depict the times at which outgoing e-mails were recorded by an e-mail
server for four separate individuals~\cite{kossinets06}, while the fifth is
synthetically generated from our model.  Based on visual inspection alone, it
is unclear how these time series should be classified~\cite{keogh05}. As we
will show, however, the synthetic time series can be easily identified in terms
of our model parameters, while the remaining users cluster into one of two
interpretable categories of human activity patterns. Having identified these
categories, we further demonstrate that 75\% of individuals remain in the same
category over the course of a two-year period, thereby reinforcing our
claim that individuals can be usefully characterized in terms of their
communication patterns.

The remainder of this manuscript is organized as follows. In section
\secref{\ref{sec:model}}, we formalize our model and show that it can be
phrased as a variant of the well-studied hidden Markov model
(HMM)~\cite{rabiner1989}. Leveraging this formulation of the model, we
implement a computationally efficient expect\-ation-maximization (EM) inference
algorithm~\cite{dempster1977} to estimate model parameters from the observed
data, which we validate with synthetic data. In \secref{\ref{sec:data_sets}} we
describe the two data sets~\cite{eckmann04,kossinets06} studied, and we present
the analysis of the inferred parameters of all users under consideration in
\secref{\ref{sec:results}}.  Open questions and future directions are discussed
in \secref{\ref{sec:discussion}}.

\section{Model}
\label{sec:model}

\begin{figure}[!tb]
\epsfig{file=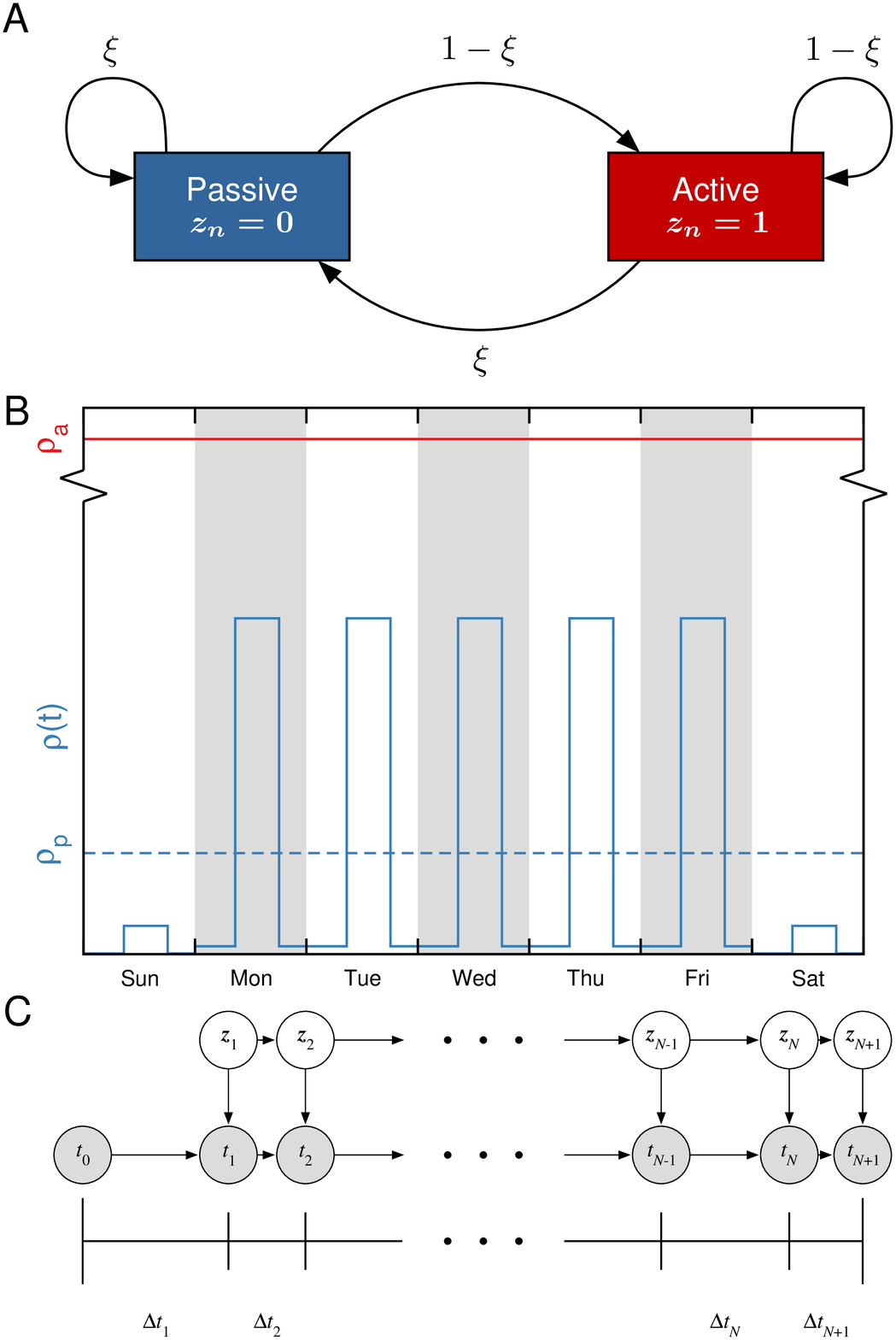,width=\columnwidth}
\vspace*{-9pt}
\caption{
A \cnhpp for human activity patterns.
{\bf \panela}, Human activity patterns are characterized by a first-order
Markov process with transition rates $\boldsymbol{\qq}$ and $\boldsymbol{
1-\qq}$ between an active state ($\boldsymbol{ \zn{\n}=1}$) and a passive state
($\boldsymbol{ \zn{\n}=0}$) of activity.
{\bf \panelb}, In the active state, e-mails are sent at a relatively high
intra-session rate $\boldsymbol{ \ratea}$ (solid red line) whereas in the
passive state, e-mails are sent at a relatively low inter-session rate
$\boldsymbol{ \ratet{\ttt}}$ (solid blue line).  On average, the rate during the
passive state is $\boldsymbol{ \ratep}$ (dashed blue line).
{\bf \panelc}, Our model can be viewed as a double-chain Markov model where the
time between consecutive events $\boldsymbol{ \deltatn{\n}}$ is drawn from a
Poisson process with rate $\boldsymbol{ \ratea}$ or rate
$\boldsymbol{ \ratet{\ttt}}$.  White (grey) circles denote missing (observed)
data.
}
\vspace*{-9pt}
\label{fig:model}
\end{figure}

The \cnhpp we pre\-sent is motivated by two key observations: first, 
individuals send e-mail during ``sessions'' of relatively high activity that
are separated by periods of inactivity during which no e-mails are
sent~\cite{malmgren08}; and second, the likelihood of commencing an active
session is modulated by daily and weekly cycles.  For convenience, we define
the start and end of a session by the first and last e-mails sent in that
session respectively.
We define an individual as ``active'' if they are in an e-mail session, where
the time between consecutive e-mails within each session is modeled as a
homogeneous Poisson process with intra-session rate $\ratea$.  Correspondingly,
we define an individual as ``passive'' if they are between e-mail sessions,
where the time between sessions is modeled as a \nhpp with inter-session rate
$\ratet{\ttt}$, which explicitly accounts for daily and weekly cycles of
activity.
Finally, we assume that after each communication an individual transitions to the
passive state with transition probability $\qq$.  In this way, our model
can be viewed as a mixture of two Poisson process whose transitions are
governed by a first-order Markov process with control parameter $\qq$
(\figref{\ref{fig:model}\panela--\panelb}).


Before proceeding, we note two features of the inter-session rate that are essential for
capturing human activity.  First, the inter-session rate $\ratet{\ttt}$ depends
on time in a periodic manner; that is, $\ratet{\ttt} = \ratet{\ttt + W}$ where
$W$, the period of the process, is defined as one week.  Second, we
parameterize the form of the inter-session rate $\ratet{\ttt}$ as
\begin{equation}
  \ratet{\ttt} = \ratep\,W\,\pdfull\,\pwfull, 
  \label{eqn:nhpp}
\end{equation}
where $\ratep$ is the average inter-session rate, $\pd$ is the
probability of starting a session during a particular hour of the day,
and $\pw$ is the probability of starting a session during a particular
day of week.  In contrast with previous work~\cite{malmgren08}, we
parameterize both $\pd$ and $\pw$ by a square pulse distribution on a
circle of circumference $\tau=24$ hours and $\tau=7$ days
respectively, where in each case the form of the square pulse is given
by
\begin{eqnarray}
  \pulse{\tauz}{\tauo}{\varepsilon} &=& \left\{
           \begin{array}{rl}
           \omega , & \quad \overline{\ttt} \in \left[\tau_0,\tau_1\right)
             \\ \varepsilon \, \omega, & \quad {\rm otherwise}
	   \end{array}
           \right. ,
\label{eqn:pulse}
\end{eqnarray}
$\omega=\left[\varepsilon\,\tau+(1-\varepsilon)(\overline{\tauo-\tauz})\right]^{-1}$
by normalization and $\overline{x}$ denotes arithmetic modulus $\tau$.  Moreover, we
refer to $\tauzd$ (and respectively $\tauzw$) as the start of the day (week),
$\tauod$ ($\tauow$) as the end of the day (week), and
$\varepsilon_d\in\left[0,1\right]$ ($\varepsilon_w$) as the modulation during
inactive periods of the day (week).  
This parameterization of the inter-session rate $\ratet{\ttt}$ therefore accounts for the
heightened tendency of users to commence an active session during the waking
hours of the day and working days of the week
(\figref{\ref{fig:model}\panelb}).
%
%
For example, if Sid works a ``nine to five'' job and he checks his
e-mails every other weekend, we might expect that he starts
sending e-mails at 09:00 ($\tauzd=9$), he stops sending e-mails at 18:00
($\tauod=18$), he is completely inactive at night ($\varepsilon_d=0$), and he
only sends e-mails Monday ($\tauzw=0$) through Friday ($\tauow=5$) as well as
every other weekend ($\varepsilon_w=0.5$).

Our general objective of characterizing individual communication behavior therefore reduces to  estimating the parameters $\bftheta=\{\paramslist\}$ for a sequence of observed
events. The main technical difficulty associated with this estimation is that we lack knowledge of session information; that is, we only
have access to event times without any record of which events
belong to which sessions, prohibiting us from directly estimating the
intra-session rate $\ratea$, the inter-session rate $\ratet{\ttt}$, or
the transition probability $\qq$. To address this issue, we consider
the assignments of e-mails to sessions as missing data, which we then
infer from the event sequences. We accomplish this task in an
efficient and scalable manner by noting that the model can be specified as a hidden Markov
model (HMM); thus expectation-maximization (EM) can be used to
simultaneously infer both the assignments of e-mails to sessions and
the parameters $\bftheta$.

To formalize these ideas, we index the observed events by
$\n=1,\ldots,\N$ and denote the time of the $\n^{\rm th}$ observation
as $\tn{\n}$.  Importantly, all of the events in this time series are
observed over a finite time interval, which, for ease of notation, we
will denote by the interval $\left[\ttt_0,\ttt_{\N+1}\right)$.  We
then introduce the latent variable $\zn{\n}$ to denote that an
individual is passively sending e-mail ($\zn{\n}=0$) between times $\tn{\n-1}$
and $\tn{\n}$ (\emph{i.e.} the $\n^{\rm th}$ event corresponds to the beginning
of a new session) or actively sending e-mail ($\zn{\n}=1$) between times
$\tn{\n-1}$ and $\tn{\n}$ (\emph{i.e.} the $\n^{\rm th}$ event is a
continuation of the current session). 
%
We refer to the sets of all observed and latent variables as
$\bft=\{\ttt_0,\ttt_1\ldots,\ttt_{\N+1}\}$ and $\bfz=\{z_1,\ldots,z_{N+1}\}$,
respectively.  Given the state of the individual $\zn{\n}$, the time
$\deltatn{\n}=\tn{\n} - \tn{\n-1}$ between two consecutive events is governed
by the corresponding rate; that is, the time of event $\tn{\n}$ depends on both
the state of the individual $\zn{\n}$ and the time of the previous event
$\tn{\n-1}$.  Whereas a standard HMM has first-order couplings between latent
variables only, our model has additional first-order couplings between
observations.  As a result of its graphical model structure
(\figref{\ref{fig:model}\panelc}), it is referred to as a double-chain HMM
(DCHMM)\footnote{Models with this structure have also been termed
  autoregressive HMMs (ARHMMs).  This term, however, often implies a specific
  form for the coupling between observations that differs from the one used
  here, so we refer to the model as a DCHMM to avoid confusion.}.
In contrast with the conventional HMM formulation
where each element in the Markov chain is separated by equally spaced
intervals in time (e.g. Ref.~\cite{scott99}), the Markov chain used
here is indexed by event number and the emissions correspond to event
times.  As such, the elapsed times between elements in the chain are
not identical.

\begin{figure}[!tb]
\epsfig{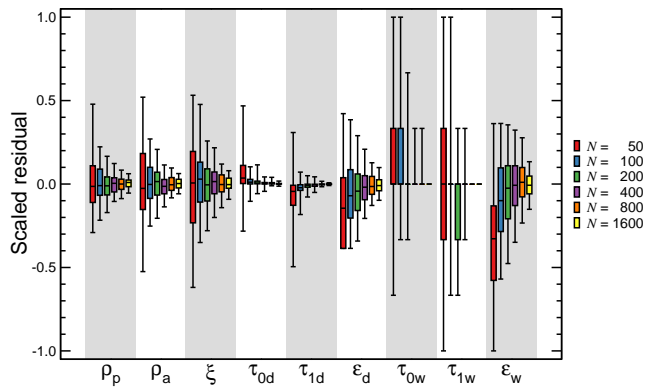}
\vspace*{-9pt}
\caption{
Validation of the EM parameter estimation procedure.  
Boxes depict the inter-quartile range and the
whiskers depict the 95\% confidence interval. 
}
\vspace*{-9pt}
\label{fig:validation}
\end{figure}

%
Having formulated the model as a DCHMM, we utilize EM to infer the model
parameters $\bftheta$ and missing session assignments $\mathbf{z}$ from the
observed event times $\mathbf{t}$.  
The EM algorithm iteratively updates the probability distributions over the
unobserved session assignments and the maximum-likelihood parameter estimates
in a two-step process that proceeds until it converges to a local optimum of
the incomplete-data likelihood.  In the E-step, the distribution over session
assignments, conditioned on the current parameter estimates, is calculated
using the forward-backward algorithm~\cite{rabiner1989}. \\ Note that despite the
coupling of successive emissions from the DCHMM, the forward-backward algorithm
can still be used for efficient calculation of the E-step~\cite{berchtold99}.
In the M-step, the parameter estimates are updated to their maximum-likelihood
values, conditioned on the current distribution over session assignments.
After the likelihood converges to a locally optimal solution, the procedure is
repeated several times (here, 25) to improve the probability of finding the
globally optimal solution.  Compared with the $\mathcal{O}(\N^2\ln\N)$ run
time required by simulated annealing inference for the complex model in
Ref.~\cite{malmgren08}, EM inference, which requires $\mathcal{O}(\N)$ time,
offers significant computational savings.  For example, EM inference for this
model takes less than a minute on a standard desktop computer for a typical
time series of 100 events over a 4-month period, whereas parameter inference
for the related model in Ref.~\cite{malmgren08} requires a few hours for the
same calculation.
%
Thus, in simplifying the original \cnhppnospace, we gain not only
interpretability, but also the ability to estimate the model for time series
with thousands of events and for data sets comprising a large number of
individuals (i.e. at least tens of thousands).

Using simulated data generated from our model, we have verified that 
our EM procedure provides an asymptotic unbiased estimate of model
parameters.  Specifically, we generated 100 synthetic time series,
each with $\N$ events from our model with randomly chosen parameters
$\bftheta_o$. We then used EM to compute the best-estimate parameters
$\widehat{\bftheta}$ for each time series and computed the scaled
residuals between the best-estimate parameters and the actual
parameter setting $\widehat{\bftheta}-\bftheta_o$ where the residuals
are scaled by the maximum residual for each parameter.  As shown in
(\figref{\ref{fig:validation}}), the scaled residuals asymptotically
approach zero for each parameter, suggesting that parameter estimates
are asymptotically unbiased (where we have confirmed that these
results are insensitive to the choice of $\bftheta_o$).

\section{Data sets}
\label{sec:data_sets}

We consider e-mail communication records from two anonymous universities where
each e-mail record comprises a sender identifier, a recipient identifier, the
size in bytes of the e-mail and a time stamp.
The first data set, collected by Eckmann, Moses and Sergi
(EMS)~\cite{eckmann04}, comprises e-mail records for 3,188 e-mail accounts at a
European university over an 83-day period.  The time stamp for each e-mail
record has a precision of one second and only e-mails sent between university
members are recorded.  These data were collected several years ago, before home
Internet access was common~\cite{eckmann08}; thus it is likely, for example,
that activity on weekends is systematically lower than for an equivalent
contemporary population.

The second data set, originally studied by Kossinets
and Watts (KW)~\cite{kossinets06}, comprises 122,133 e-mail accounts
record\-ed from a US university's e-mail server over a 6-semester
(2-year) period.  Each e-mail record has a precision of one minute.
We note that the KW data was collected a few years later than EMS,
after it had become common for students to have access to their
university e-mail accounts on weekends.  Moreover, in contrast with the
EMS data set, the KW data includes e-mails within and outside of the
university, and also includes additional metadata for each university
member including, for instance, each individual's affiliation with the
university.
Finally, the KW data includes not only students, but also faculty
and staff whose behavior and routines we might expect to be
systematically different from that of students.  To make our
comparisons between the EMS and KW data set as relevant as possible,
therefore, we restrict our analysis of the KW data set to
undergraduate and graduate students during the four non-summer
semesters, denoted as semesters 1, 2, 4, and 5.

In addition to their university e-mail account, individuals at each university
presumably communicate through several other modes of communication, such as
face-to-face conversation, telephone, text messaging, instant messaging, and
external e-mail accounts.  As a result, some individuals may favor other modes
of communication over their university e-mail account, possibly to the point of
not using it at all; indeed, this seems to have been the case. EMS User 1962,
for example, only sent 5 e-mails while receiving 2,284 e-mails and KW User
3069337 only sent 1 e-mail while receiving 26,230 e-mails.  To restrict our
study to users who use their university account as a primary mode of electronic
communication, therefore, we limit our analysis to individuals who sent, at
minimum, approximately 1 e-mail every 2 days. In addition, we consider only
e-mail accounts that are likely not listservs, and we collapse all recipient
lists of e-mails sent within 5 sec.~(1 min.) into one e-mail to be consistent
with previous work~\cite{malmgren08}. As a result of these steps, the data we
consider comprise 404 e-mail accounts from the EMS data set which sent at least
41 e-mails over 83 days and 6,164 student e-mail accounts from the KW data set
which sent at least 50 e-mails during each of the four non-summer semesters.

\section{Results}
\label{sec:results}


We now turn our attention to assessing the
feasibility of characterizing user behavior with the parameterization of a
simple mechanistic model.  First, we compute the best-estimate parameters of
all users under consideration for the EMS data set and each of the four
non-summer semesters of the KW data set.  Since the EMS data set is from a
European university in an era before widespread home Internet access whereas
the KW data set is from an American university some years later, one might
expect that the parameter estimates for the two data sets would be markedly
different.  As \figref{\ref{fig:p_theta}\paneld--\paneli} shows, the EMS
parameter estimate distributions do in fact differ from the KW distributions
for the model parameters which describe daily and weekly fluctuations,
presumably due to the strong influence of work cycles on the EMS users. Yet as
\figref{\ref{fig:p_theta}} shows, the parameter estimate distributions for the
two data sets are otherwise surprisingly similar; in particular, the users in
the EMS and KW data sets have similar rates of starting sessions, rates of
sending e-mails during active intervals, and probabilities of terminating an
active interval (\figref{\ref{fig:p_theta}\panela--\panelc}).

In addition to comparing the two universities, we have also considered the
similarities and differences between the four non-summer semesters of the KW
data set.  One might reasonably expect that substantial changes would manifest
themselves from semester to semester as students enter and leave the
university, change their course schedules, and become increasingly reliant on
e-mail communication. Consistent with this intuition, we identify three minor
systematic changes over the four semesters: the log-mean inter-session rate
$\langle\log\ratei\rangle$ increases; the log-mean intra-session rate
$\langle\log\ratep\rangle$ increases; and the mean probability
$\langle\qq\rangle$ of terminating an e-mail session decreases.  These
nonstationarities are indicative of a population that utilizes more e-mail
sessions and sending more e-mails per session; however it is unclear whether
this increased e-mail usage is due to endogenous factors (\emph{e.g.}
increasingly selecting e-mail as a primary mode of communication), exogenous
factors (\emph{e.g.} increasingly responding to e-mails from others), or a
selection bias from considering only those users that sent at least 50 e-mails
in each of the four non-summer semesters. Aside from these minor changes,
we find that the KW parameter estimate distributions are markedly
similar across successive semesters.

\begin{figure*}[!tb]
\epsfig{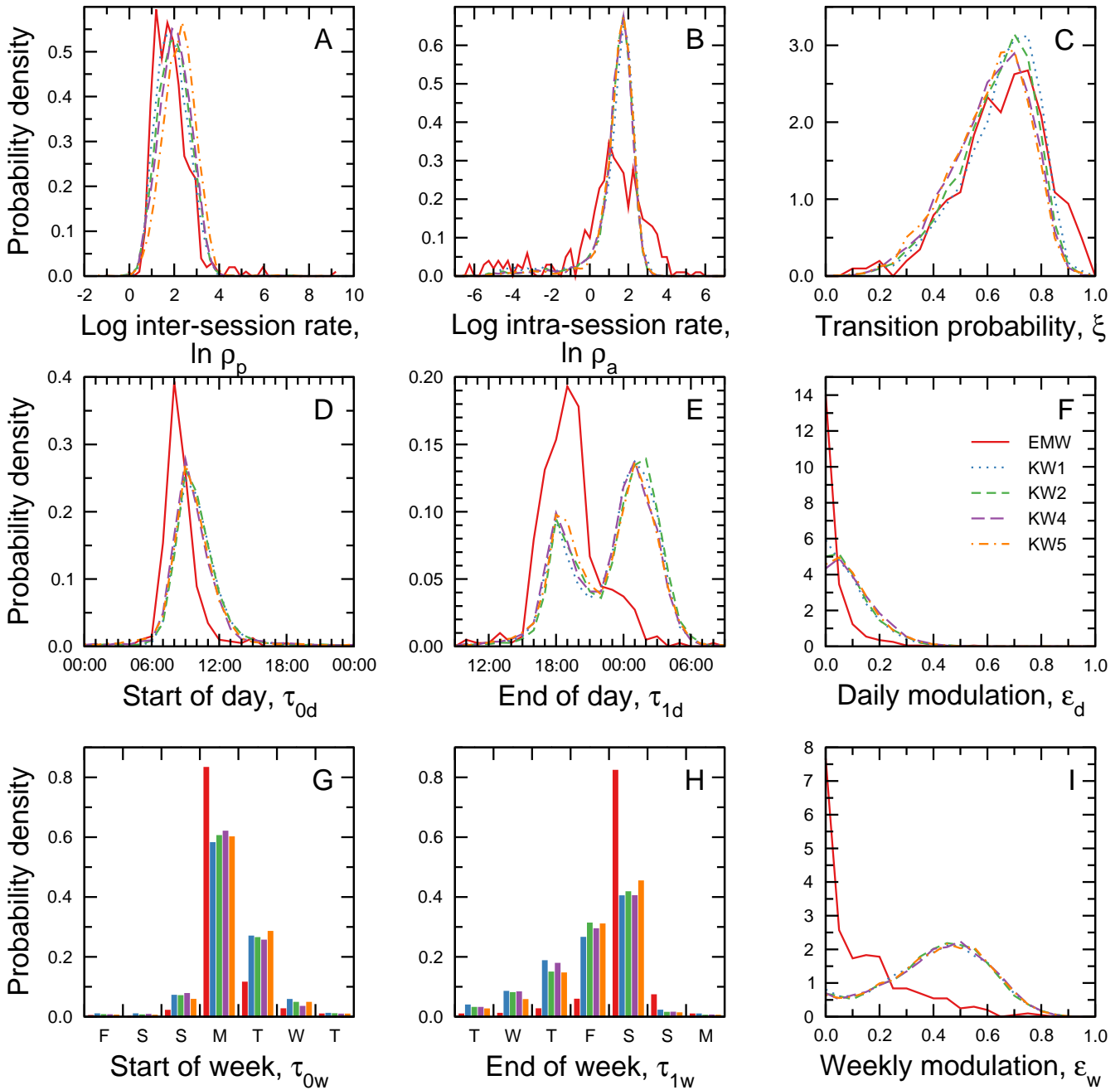}
\caption{
Best-estimate parameter distributions for the users in the EMS data set and KW
data set during semesters 1, 2, 4, and 5.  Differences and similarities of
these distributions are discussed in the text.
}
\label{fig:p_theta}
\end{figure*}


The overall similarity of these distributions, both across different
universities and also across different semesters of the same university,
therefore suggests that our model does indeed capture salient features of human
communication patterns that are common across different times and contexts.
From the perspective of using communication data to characterize individual
behavior, in other words, these findings are encouraging; however, they also
raise an important question: does the heterogeneity observed in the parameter
estimate distributions arise from ``typical'' individuals who approximately
change over time in the same manner, or does it instead indicate the presence
of heterogeneous individuals whose behavior is relatively stable over time?


\begin{figure*}[!tb]
\epsfig{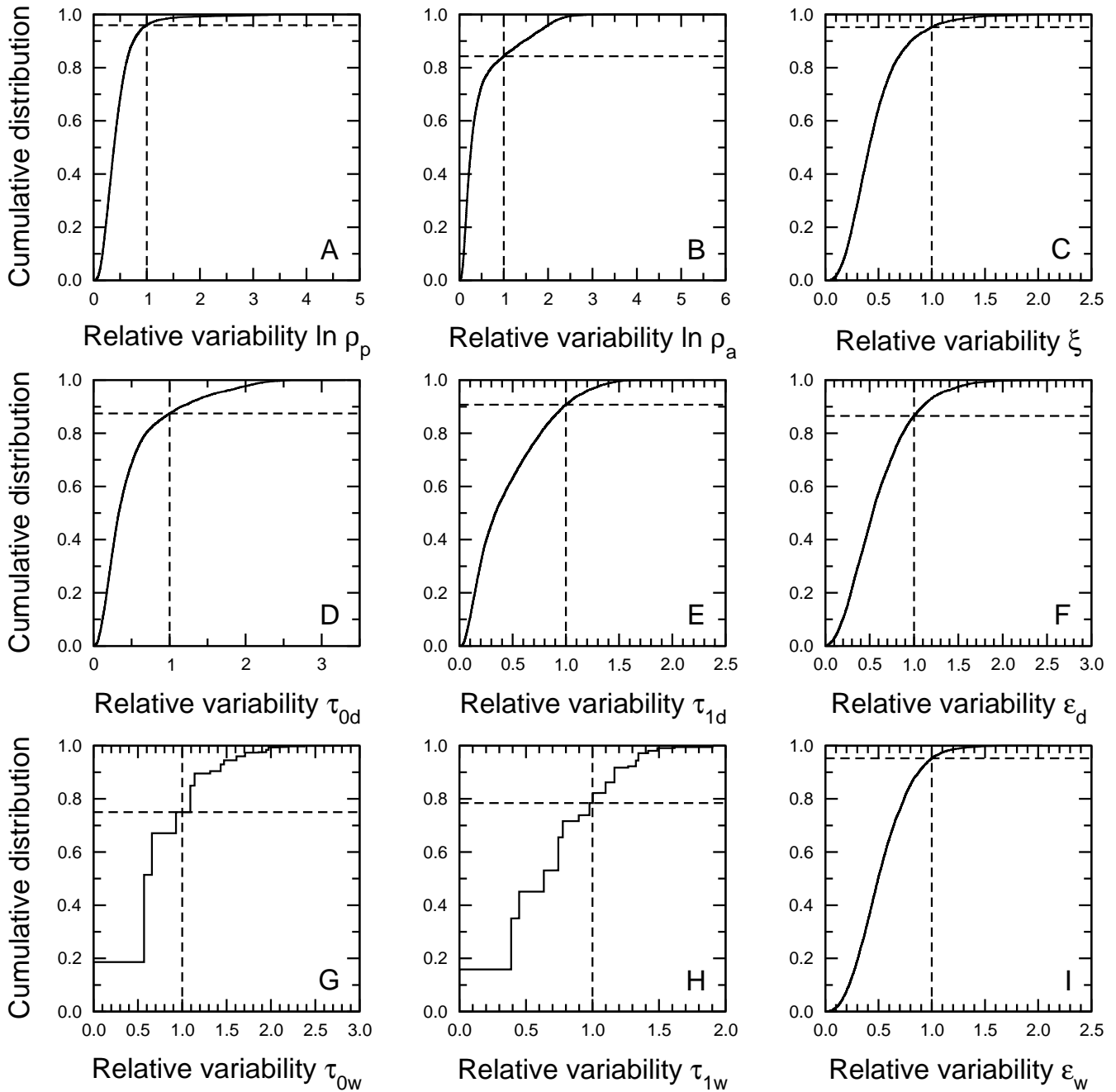}
\caption{ 
Complementary cumulative distribution of relative individual
variability, the ratio between the within-individual and between-individual
variability. Dashed lines are a guide for the eye, indicating the fraction of
individuals whose within-individual variability is less than
the between-individual variability.
}
\label{fig:var_theta}
\end{figure*}

To address this question, we next examine the inter-sem\-ester fluctuations of
the parameter estimates for users within the KW data set and compare these
``within-individual'' variations (i.e. over time) with the
``between-individual'' variations (i.e. across the population) of all
individuals for any given semester.  Specifically, for each of the nine
parameters, we define within-individual variability as the standard deviation
of the relevant parameter estimated for each user over the four
semesters $s=1,2,4,5$. Correspondingly, we define the between-individual
variability for each parameter as the standard deviation of the relevant
distribution averaged over the four semesters. Within-individual variability
therefore represents the characteristic scale by which an individual ``moves''
over time in parameter space, while between-individual variability sets a
natural length scale against which the former distance can be measured. In
\figref{\ref{fig:var_theta}}, we plot the complementary cumulative distribution
function of the relative variability---the ratio of the within- and
between-individual variability---and find that for almost all parameters, more
than 80\% of individuals vary less over time than they do across the population
for any single semester.
This finding is significant for two reasons. First, it demonstrates that human
behavior cannot be well represented by a single, universal set of canonical
rules, as has been claimed previously~\cite{vazquez06}; rather, individuals
have distinct behaviors that persist over extended periods of time relative to
the variations between individuals.  And second, it suggests that it may be
able to exploit these persistent behaviors to classify individuals into
``types'' according to their activity patterns.


To further explore this possibility, we perform a very simple clustering of
users in a sub-space of our model's parameter space.  Specifically, we examine
the joint two-dimensional distribution of parameters $\tauzd$ and $\tauod$,
which capture when users start and end their days, respectively.  This
cross-section of our parameter space is interesting in that obtaining the same
information using standard time series clustering techniques would be very
difficult.  We fit a standard two-component Gaussian mixture model using EM to
the joint distribution~\cite{bishop07} and find that, users can be placed
broadly into two clusters, as shown in
\figref{\ref{fig:theta_classification}\panela}. Moreover, the two clusters have
clear interpretations arising from the intuitive nature of the model
parameters: the first cluster (blue dots) comprises ``day laborers'' (like our
hypothetical example of Sid from earlier) who send e-mail during working hours
from 09:00 to 18:00, whereas the second cluster (red crosses) comprises
``e-mailaholics'' who send e-mail throughout their waking hours from 09:00 to
01:00.

Once noted, the distinction between ``day laborers'' and ``e-mailaholics''
seems clear, even obvious. Superposing the location of the five users whose time series were
depicted in \figref{\ref{fig:time_series}}, however, we suggest that it is not at all clear which of the five time series are more similar to one another or what criteria should be
used to separate them. Nevertheless, both questions are immediately answered
by \figref{\ref{fig:theta_classification}\panela}; that is, users B and
C are clearly ``day laborers''), whereas D and E
are clearly ``e-mailaholics''.  User A, meanwhile,
does not appear to belong to either cluster---an informative
exception, as the data corresponding to user A was in fact the time
series that was artificially generated, and therefore does not
correspond to any ``real'' user at all.  Although in this instance we
have hand-picked model parameters outside of the typical range of
human activity patterns, this exercise nonetheless demonstrates that
it is be possible to detect outlier behavior
that may not be easily detectable
by other methods.  For instance, one might imagine that spammers,
like our synthetic user, might disguise their identity by sending
e-mails at a rate that mimics a normal pattern, but may give themselves
away by starting their ``day'' at an abnormal time, as indeed we see
in \figref{\ref{fig:theta_classification}\panela}.


Having identified these two distinct clusters of behavior, we can now restate
our earlier question about the temporal stability of user behavior more
concretely; that is, do individuals retain the same type of behavior from
semester to semester or do they switch their behavior depending on the
circumstances every semester?  We emphasize that our previous results,
illustrated in \figref{\ref{fig:var_theta}}, do not answer this question on
their own, as even if individuals tend to fluctuate less over time than the
cross-sectional variability of the population, they may still fluctuate enough
to switch clusters. As \figref{\ref{fig:theta_classification}\panelb}
indicates, however, the vast majority (77\%) of individuals are indeed
``stable" in the sense that if they start in a given cluster, they will remain
in that cluster the next semester---in fact, nearly as many (74\%) remain in
the same cluster for the entire two year period.  In other words, most
individuals appear to retain their routines over extended periods of time, in
spite of their changing circumstances and course schedules, implying that our
method of characterizing them may be broadly applicable.  Nevertheless, we note
that a sizable minority---roughly 25\%---do change clusters over the duration
of the KW data.  This minority appears to be randomly scattered throughout
parameter space, raising interesting and unresolved questions about classifying
individual behavioral patterns.

\begin{figure}[!tb]
\begin{center}
\epsfig{file=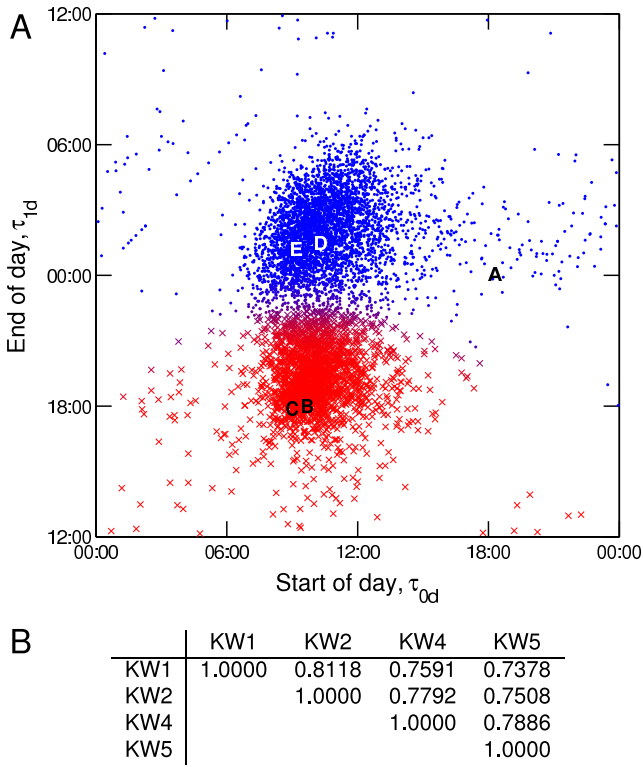,width=\columnwidth}
\end{center}
\vspace*{-9pt}
\caption{
Stable clustering of users in parameter space.  
{\bf \panela}, Two-dimensional Gaussian mixture model fit to the joint
distribution of $\boldsymbol{\tauzd}$ and $\boldsymbol{\tauod}$ for semester 1.
{\bf \panelb}, Stability of cluster assignments where each entry in the table
indicates the fraction of users who are found in the same cluster between the
two corresponding semesters.
}
\label{fig:theta_classification}
\vspace*{-9pt}
\end{figure}

\section{Discussion}
\label{sec:discussion}

In this manuscript, we have introduced a simplified version of the
\cnhpp \cite{malmgren08} that is amenable to computationally efficient
inference techniques suitable for the large scale characterization of human
communication patterns.  We have applied this model to two e-mail data sets
collected from different universities during different years, and we have
presented four main findings.  First, we find that the distributions of
parameter estimates for our model are generally similar across data
sets---remarkably so given the difference in circumstances under which the data
was collected---and stable in time. Second, we find that the parameter
estimates for over 80\% of individuals in the KW data vary less over the
course of four semesters than individuals vary among each other during any one
semester.  Third, we show that the population can be sensibly partitioned into
two distinct clusters that correspond to easily interpretable differences in
individual behavior. And finally, we find that our earlier characterization of
individuals as ``stable'' over time also applies to these clusters: roughly
75\% of users remain in the same cluster over a two-year interval.

As suggested by our analysis of user clusters, inferred model
parameters can be used as features for the tasks of clustering,
classification, outlier detection, or change-point
analysis~\cite{scott99} in large-scale online systems. Although initial
parameter inference takes on the order of minutes for most users, the
developed algorithms can easily be parallelized on a per-user basis,
and on-line HMM inference algorithms~\cite{krishnamurthy93} can be
leveraged for inexpensive updates to model parameters as new data are
recorded. Future directions for work include evaluation of the model
parameters as features for the task of classifying known malicious
users.

Although we have focused on a cascading non-homogen\-eous Poisson process in this
manuscript, we conclude by reiterating that our principal motivation is to
leverage the rapidly increasing volume of communication data---and even more
generally, activity data---to characterize individuals. From this perspective,
the details of the particular model are of secondary importance, and we
anticipate that other models may be more appropriate in different contexts or
for different sorts of communication data.  The emphasis here is that, by using
a simple mechanistic model that captures the salient features of human
activity, it is possible to meaningfully characterize human activity and to use
this characterization as another attribute of sociological study.

We do not claim, however, that individual attributes extracted from
communication or activity data are more or less informative than demographic or
network-based attributes.  Depending on the question of interest and the
available data, one of these approaches may be more suitable than others, and
further work is needed to address this issue, possibly comparing the predictive
or explanatory power of the various methods directly.  Alternatively, one might
also consider hybrid approaches that make use of more than one kind of data,
analogous to ``social targeting'' methods~\cite{hill06} which leverage network
data to substitute for missing attribute data, based on the so-called
``homophily principle''~\cite{mcpherson87} that friends are more similar than
strangers. In the same vein, one might leverage observable communication
patterns to predict unobservable categorical data, like affiliation or status.
For example, do faculty communicate differently than students, or do highly
central nodes also distinguish themselves in their communication patterns?
Regardless of the specific applications, we anticipate that with the increasing
availability of communication data, the dynamics of communication will serve as
a useful attribute in understanding the behavior of individuals.

\section{Acknowledgements}

We thank D.B.~Stouffer, W.~Mason, S.~Goel, S.~Suri, R.~Guimer\`a,
M.~Sales-Pardo, and M.J.~Stringer for insightful comments and suggestions.
Special thanks to G.~Kossinets and J.P.~Eckmann for their help with the data.
L.A.N.A. gratefully acknowledges the support of NSF award SBE \\0624318 and of
the W.~M.~Keck Foundation.  Figs.~1--5 were generated with PyGrace
(http://pygrace.sourceforge.net) \\with color schemes from
http://colorbrewer.org.

\bibliographystyle{abbrv}
\bibliography{./References/ref-database,./references}

\end{document}